# OpenAlex2Pajek – an R Package for converting OpenAlex bibliographic data into Pajek networks[1]

**Vladimir Batagelj**
IMFM Ljubljana and IAM UP Koper, Slovenia
e-mail: `vladimir.batagelj@fmf.uni-lj.si`
ORCID: 0000-0002-0240-9446

For analysis of bibliographic data, we can obtain from bibliographic databases the corresponding collection of bibliographic networks. Recently OpenAlex, a new open-access bibliographic database, became available. We present OpenAlex2Pajek, an R package for converting OpenAlex data into a collection of Pajek's networks. For an illustration, we created a temporal weighted network describing the co-authorship between world countries for years from 1990 to 2023. We present some analyses of this network.

## Introduction

"OpenAlex is a fully open catalog of the global research system. It's named after the ancient Library of Alexandria and made by the nonprofit OurResearch" (Priem et al., 2022, OpenAlex, 2024, OurResearch, 2024). OpenAlex launched in January 2022 with a free API and data snapshot. It is considered an alternative to the Microsoft Academic Graph, which retired on Dec 31, 2021 (Chawla, 2022). OpenAlex is based on 7 types of units (entities): **W**(ork), **A**(uthor), **S**(ource), **I**(nstitution), **C**(oncept), **P**(ublisher), or **F**(under). It solves some important questions for the analysis of bibliographic data:
- identification of bibliographic units (IDs, disambiguation)
- free access (share derived data, download to your machine)
- improving content through user participation (submit a request)

OpenAlex opened a space for the development of higher-level bibliographic services using bibliographic data analysis to advise the user. For example: a selection of reviewers, a selection of a journal to publish an article, an analysis of publication activity of a research group or institution, etc.

We developed in R a package of functions `OpenAlex2Pajek` (OpenAlex2Pajek, 2024) for constructing bibliographic networks from selected bibliographic data in OpenAlex. Currently, `OpenAlex2Pajek` contains three main functions `OpenAlex2PajekCite`, `OpenAlex2PajekAll`, and `coAuthorship`.

## Creating the collection of bibliographic networks

In 2007 we developed in Python a program `WOS2Pajek` for constructing bibliographic networks from selected bibliographic data from WOS (Web Of Science). `OpenAlex2Pajek` is based on experiences gained using this program.

We split the process of creating the collection of bibliographic networks into two parts:

---

[1] Draft version with vector graphics color figures of the paper published in Collnet 2024 proceedings ISBN 978-93-86578-65-5.

- determining the set *W* of relevant works using the ***saturation approach*** (Batagelj et al., 2017, page 506],
- creation of the network collection for the works from *W*.

The set *W* is determined iteratively using the function `OpenAlex2PajekCite` and the collection is finally created using the function `OpenAlex2PajekAll`.

After each run of the function `OpenAlex2PajekCite` we read the last version of the citation network into Pajek (De Nooy et al., 2018) and apply macro `expNodes` to it. It produces a vector of expansion nodes. Using the vector-Info button in Pajek we get a list of works with the largest input degree. We select an appropriate threshold and extract (select and copy) the upper part of the table into `TextPad`. In `TextPad`, we remove other columns and save the list of works as a `CSV` file. Using the function `joinLists` we combine the old list of works with the new one and save it for the next step of the saturation procedure.

The collection contains the citation network **Cite** and two-mode networks: authorship **WA**, sources **WJ**, keywords **WK**, countries **WC**, and work properties: publication **year**, **type** of publication, the **language** of publication, **cited** by count, countries **distinct** count, and **referenced** works.

Mark Batagelj (2024) used this approach to make a collection of networks on the topic of handball. In some cases, such as all works of researchers from a selected institution, the saturation phase is not needed.

## Co-authorship between world countries

From OpenAlex we can collect the data about the co-authorship between world countries. To get a selected country, for example, SI, collaboration list we use the query

https://api.openalex.org/works?filter=authorships.countries:SI&group-by=authorships.countries

We developed a function `coAuthorship` that creates a temporal network describing the co-authorship between world countries in selected time periods. It turned out that OpenAlex is using the current ISO 3166-1 alpha-2 (2024) two-letter country codes to represent countries, dependent territories, and special areas of geographical interest. It doesn't consider ex-countries such as SU (Soviet Union) or YU (Yugoslavia) – such allocations are transformed into the corresponding current countries. Another problem in creating the co-authorship network between world countries is that the above query returns information about up to 200 most collaborative countries. The problem is resolved by considering the symmetry of the co-authorship data.

*Table 1: Example*

| paper | countries | | | AU | DE | ES | IT | SI | US |
|---|---|---|---|---|---|---|---|---|---|
| W2001947224 | SI, US, SI | | AU | 1 | | | | 1 | |
| W2021064255 | ES, SI, ES, ES | | DE | | 1 | | 1 | 1 | |
| W1984191816 | AU, SI, AU, AU, AU, SI | | ES | | | 2 | | 2 | |
| W2096814473 | SI, DE, IT, IT, IT, IT | | IT | | 1 | | 1 | 1 | |
| W2514227811 | ES, ES, ES, SI, ES | | SI | 1 | 1 | 2 | 1 | 6 | 2 |
| W1981385379 | US, SI, SI | | US | | | | | 2 | 2 |

The co-authorship between countries can be measured in different ways (Batagelj, V, 2024). What exactly the information obtained from OpenAlex is measuring? To answer this question, we applied the query to 6 papers listed on the left side of Table 1. On the right side, we have the corresponding co-authorship matrix **Co** = [ Co[*a,b*] ]. In the co-authorship matrix **Co,** non-existing links are represented with the value NA. We see: let *W* be the set of works with co-authors from at least 2 different countries. $W_a \subseteq W$ is the set of works with an author from the country *a*. For $a \neq b$, Co[*a,b*] = $|W_a \cap W_b|$ – the number of works with co-authors from countries *a* and *b*; and Co[*a,a*] = $|W_a|$ – the number of works co-authored by authors from country *a*.

Let us denote the row sum $R(a)$ = woutdeg(a) = $\sum_b Co[a, b]$ .

Using the function `coAuthorship` we created the sequence of co-authorship networks for each year from 1990 till 2023. They are available at OpenAlex/countries (2024).

*Table 2: Nodes with the highest weighted degree core levels in the year 1990*

| i | c | t | i | c | t | i | c | t |
|---|---|---|---|---|---|---|---|---|
| 1 | US | 9791 | 16 | IN | 4124 | 31 | KR | 1736 |
| 2 | GB | 9791 | 17 | RU | 3950 | 32 | ZA | 1656 |
| 3 | CA | 9791 | 18 | PL | 3818 | 33 | SK | 1532 |
| 4 | JP | 9720 | 19 | DK | 3284 | 34 | BG | 1338 |
| 5 | DE | 9720 | 20 | AT | 2966 | 35 | PT | 1326 |
| 6 | FR | 9720 | 21 | CZ | 2894 | 36 | AR | 1202 |
| 7 | IT | 7394 | 22 | BR | 2446 | 37 | EG | 1130 |
| 8 | CH | 6884 | 23 | HU | 2424 | 38 | HK | 888 |
| 9 | NL | 6884 | 24 | NO | 2406 | 39 | CL | 810 |
| 10 | AU | 6614 | 25 | FI | 2406 | 40 | TH | 768 |
| 11 | SE | 4880 | 26 | MX | 2388 | 41 | SG | 756 |
| 12 | IL | 4818 | 27 | TW | 2308 | 42 | HR | 630 |
| 13 | CN | 4682 | 28 | GR | 1878 | 43 | SI | 630 |
| 14 | BE | 4682 | 29 | IE | 1816 | 44 | PK | 630 |
| 15 | ES | 4598 | 30 | NZ | 1756 | 45 | RS | 588 |

*Table 3: Nodes with the highest weighted degree core levels in the year 2023*

| i | c | t | i | c | t | i | c | t |
|---|---|---|---|---|---|---|---|---|
| 1 | AU | 192486 | 16 | BR | 124582 | 31 | HK | 84360 |
| 2 | US | 192486 | 17 | DK | 113534 | 32 | EG | 82306 |
| 3 | GB | 192486 | 18 | KR | 109702 | 33 | ZA | 79442 |
| 4 | IT | 192486 | 19 | AT | 107734 | 34 | GR | 78064 |
| 5 | CN | 192486 | 20 | NO | 105640 | 35 | IE | 77326 |
| 6 | DE | 192486 | 21 | PT | 101980 | 36 | IR | 77278 |
| 7 | CA | 192486 | 22 | PL | 97686 | 37 | TW | 75296 |
| 8 | FR | 192486 | 23 | SA | 95528 | 38 | MY | 71816 |
| 9 | ES | 182268 | 24 | PK | 88758 | 39 | MX | 69732 |
| 10 | NL | 178488 | 25 | CZ | 84430 | 40 | AR | 66660 |
| 11 | CH | 171110 | 26 | TR | 84430 | 41 | CL | 63632 |
| 12 | JP | 151956 | 27 | IL | 84430 | 42 | AE | 57136 |
| 13 | IN | 151956 | 28 | RU | 84430 | 43 | ID | 57136 |
| 14 | BE | 136554 | 29 | FI | 84430 | 44 | NZ | 56120 |
| 15 | SE | 134726 | 30 | SG | 84360 | 45 | HU | 50820 |

To get insight into the structure of a large network we can reduce it to its skeleton by removing less important links and/or nodes (Batagelj et al., 2014).

- Most often the ***spanning tree***, ***link cut***, or ***node cut*** are used.
- In the ***closest k-neighbor skeleton*** for each node, only *k* of the largest incident links are preserved. The resulting skeleton is invariant for monotonic transformations of weights.
- The ***Pathfinder*** algorithm was proposed in the 1980s by Schvaneveldt et al. (1988). It removes from the network with a dissimilarity weight all links that do not satisfy the triangle inequality – if a shorter path exists that connects the link's end nodes then the link is removed.
- Cores are a very efficient tool to determine the most cohesive (active) subnetworks (Batagelj & Zaveršnik, 2011). The subset of nodes $C \subseteq V$ induces a ***weighted degree*** (or $P_s$) ***core*** at level *t* if for all $v \in C$ it holds $\mathrm{wdeg}_C(v) \geq t$, and *C* is the maximum such subset. The cores are nested.

To identify the most collaborative groups of countries we applied the weighted cores procedure to the co-authorship networks for the years 1990 and 2023. The results are presented in Table 2 and Table 3. The main core in the year 1990 at level 9791 consists of US, GB, and CA. Authors from each of these three countries co-authored with the authors from the other two countries at least 9791 works. Expanding the main core with JP, DE, and FR we get the core at level 9720 -- authors from each of the core countries co-authored with the authors from other core countries at least 9720 works. Etc. We notice a huge increase in the number of (joint) publications per year 192486/9791 = 19.66. In 1990 the top 45 countries contained large or developed countries and (not small) European countries. The membership and also the ordering of countries in both tables didn't change much. In 2023, CN entered the main core and JP moved to a lower position; some medium-sized East European countries SK, BG, HR, SI, RS, and TH left the list and were replaced by larger developing countries SA, TR, IR, MY, ID, and AE.

A simple spanning skeleton that contains all network nodes is the 1-neighbors skeleton – for each node only its strongest link is preserved. The resulting directed network is forest-like. In an analysis of weighted networks, the 1-neighbor skeleton is often used to get an overall picture of the network's basic structure. In Figure 1 the 1-neighbor skeletons for years 1990, 1995, 2000, 2010, 2015, and 2020 are presented. We see that the number of isolated nodes (countries not collaborating with other countries) is decreasing. In all analyzed years the US has a leading (hub) position. Nontrivial connected components in the 1-neighbors skeletons are (usually) directed trees with a pair of nodes linked in both directions with the largest weight in the tree – these two arcs are usually replaced by an edge (undirected link). In the years 1990, 1995, 2000, and 2010 the edge in the main component links US and GB but in the years 2015 and 2020 GB is replaced by CN. In 1990, stronger secondary hubs were GB, FR, RU, JP, and DE. In the following years, some other countries SE, ES, AU, CN, BR, ZA, and IN (BRICS) became secondary hubs attracting previously non-collaborating countries or geographically or linguistically close countries. In the year 2020, we have two interesting small components: some Arabic countries SD, LY, BH, and YE linked to SA and EG, and some Muslim countries BN, IQ, and AZ linked to MY and ID.

The standard graph-based visualization of a dense network with more than 10 nodes is unreadable. A much better option for networks of moderate size (some hundreds of nodes) is the matrix representation. Another problem is the large range of the weights (for example 1:69440 in 2023) and their distribution. This problem is usually solved using some order-preserving (monotonically increasing) transformation of weights such as $w' = \sqrt{w}$ or $w' = \log w$. In the following, we used the transformation $w' = \log_2 w$ which reduces the range in our example to 0:16.0835 . An important parameter of the matrix representation is the ordering of the nodes (matrix rows and columns). Some orderings produce blocks in the matrix representation revealing the internal structure of the network. For subsets of nodes R and C, we denote by (R, C) the block (submatrix) determined by rows R and columns C.

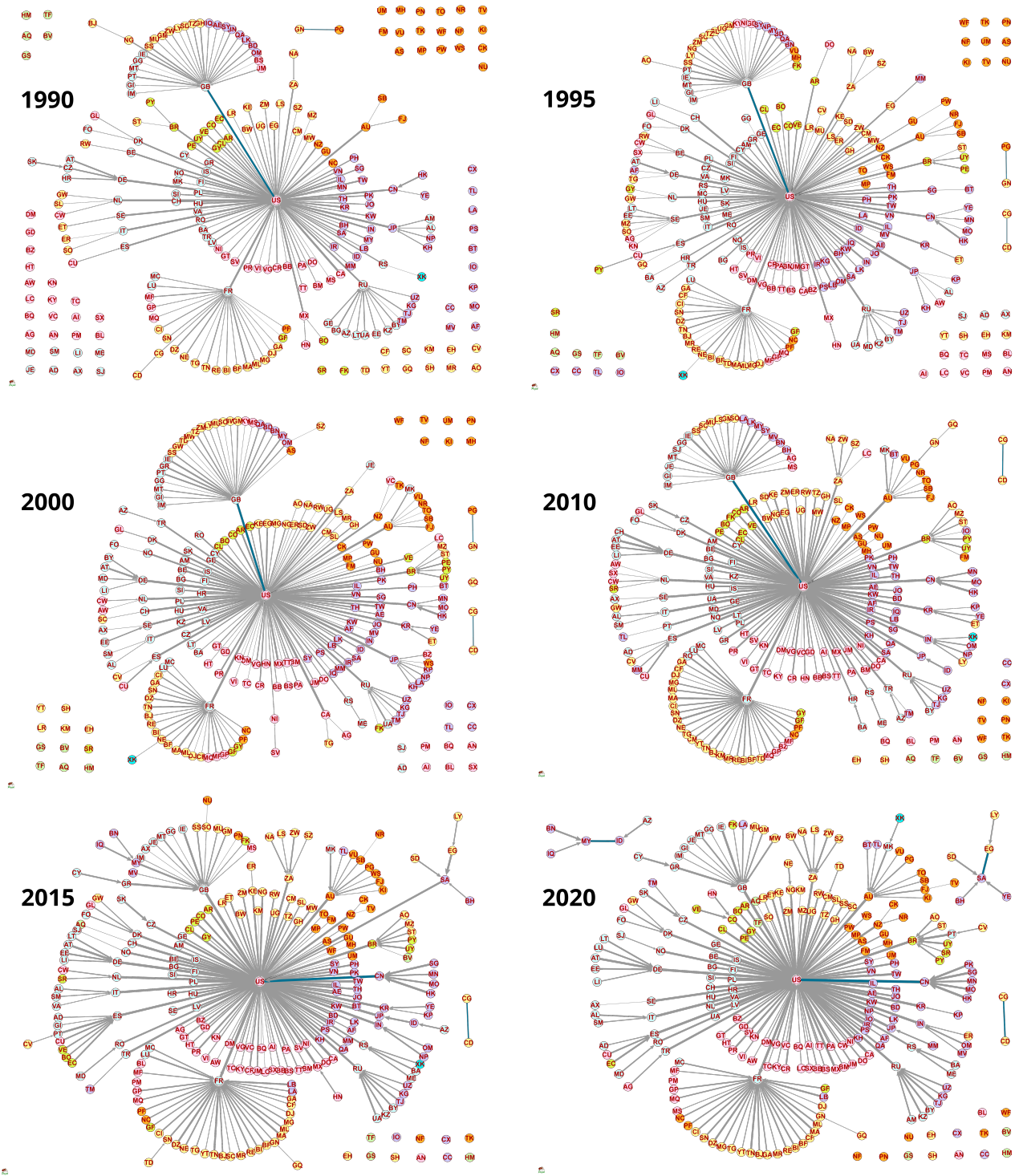

*Figure 1: 1-neighbor networks*

To find an interesting ordering is a task of blockmodeling (Doreian et al., 2004). An approach is to use the hierarchical clustering of the co-authorship network/matrix **Co**. To do this we have to define a dissimilarity $D[a,b]$ between nodes a and b in a co-authorship network. When computing the dissimilarity $D[a,b]$ it is important to use a corrected dissimilarity. We selected the corrected Euclidean distance (Doreian et al., 2004, p. 181)

$$D[a,b] = \sqrt{(Co[a,b] - Co[b,a])^2 + (Co[a,a] - Co[b,b])^2 + \sum_{c: c \neq a, c \neq b} (Co[a,c] - Co[b,c])^2}$$

For clustering co-authorship networks, we transformed the weights using $w' = \log_2 w$ – a balance between the structure (links) and weights. It is also convenient for visualization.

In the matrix representation of the co-authorship network for the year 2023 a yellow cell represents a no-link (absence of co-authorship) and a grey cell the intensity of co-authorship – the darker the cell stronger the collaboration. We identified the following clusters, see Figure 2:

$C_1$ = { GB, US, AU, CA, CN, IN, FR, ES, DE, IT, BR, BE, NL, CH, AT },
$C_2$ = { FI, IE, GR, IL, UA, HU, RO, CZ, PL, RU, TR, JP, KR, PT, DK, NO, SE },
$C_3$ = { EC, PE, AR, CL, CO, MX, BD, TH, PH, VN, HK, NZ, SG, TW, NG, ZA, ID, MY, IR, AE, PK, EG, SA },
$C_4$ = { LV, EE, LT, CY, RS, BG, SK, HR, SI },
$C_5$ = { ET, GH, KE, CM, TZ, UG },
$C_6$ = { MA, DZ, TN, KW, OM, IQ, QA, JO, LB },
$C_7$ = { GN, GM, SL, BI, MG, GA, NE, TG, CG, CD, BJ, CI, ML, BF, SN, BW, NA, MZ, RW, MW, ZM, ZW },
$C_8$ = { HT, NI, DO, SV, HN, CR, UY, VE, PR, BO, PA, CU, GT, PY },
$C_9$ = { NP, LK, KH, MN, MO, BN, MM, AF, SS, BH, PS, LY, SY, SD, YE },
$C_{10}$ = { KG, TJ, MD, XK, ME, AZ, BY, AM, GE, KZ, UZ, MK, AL, BA, LU, IS, MT },
$C_{11}$ = { JM, TT, FJ, PG, MU, SO, LA, BT, MV, BB, GD, AG, CW, KN, RE, GF, NC, PF, GP, MQ, GL, AD, FO, BS, SC, LI, MC, MR, CF, TD, LS, SZ, GY, LR, AO, GW, CV, ST, DM, BZ, LC, GI, FK, BM, SJ, SR, KY, VI, TL, WS, GU, PW },
$C_{12}$ = { KP, TM, ER, AW, SX, DJ, YT, GQ, JE, GG, IM, SM, MF, AQ, VA, CK, TO, VU, NR, KI, SB, MP, AS, FM, MS, VG, KM, MH, UM, TC, AI, VC},
$C_{13}$ = { EH, WF, TV, GS, PM, SH, BL, PN, NF, NU, AN, HM, TF, CC, BV, AX, BQ, IO, CX, TK }.

From the matrix representation in Figure 2, we first observe the ***core-periphery*** structure of the co-authorship network with the core $C_1$-$C_6$, semi-periphery $C_7$-$C_{10}$, and periphery $C_{11}$-$C_{13}$. The core countries are collaborating with each other and with many other countries. The semi-periphery countries are collaborating with most of the core countries and only some of the periphery countries. The clusters in the semi-periphery are collaborating internally but there is little collaboration between different clusters. The periphery countries are collaborating mostly with the core countries and with some semi-periphery countries. There is almost no collaboration between the periphery countries – with some exceptions such as {BB, GD, AG, CW, KN} (Lesser Antilles) and {RE, GF, NC, PF, GP, MQ} (French islands). The cluster $C_{13}$ consists of (almost) inactive countries.
Very dark cells, such as (CN, HK), (CN, MO), (HK, MO), (RU, TJ), and (UA, MZ), indicate strong collaboration between these countries. The same holds for darker rectangles in the picture, such as ({ZA, NG}, $C_5$ ) (inside Africa), ({MY, IR, AE, PK, EG, SA}, $C_6$ ) (leading Muslim countries with a group of Arabic countries), ({BH, PS, LY, SY, SD, YE}, $C_6$ ) (two groups of Arabic countries), (FR, {RE, GF, NC, PF, GP, MQ}) (France and French islands), etc.
Inspecting the row/column of a selected country we get insight into its collaboration with other countries.

The intensity of co-authorship strongly depends on the (population) size of both countries. To make countries comparable some ***normalization***s are used, such as (Matveeva et al., 2023)
- Stochastic  $M[a,b] = Co[a,b] / R(a)$
- Jaccard  $J[a,b] = Co[a,b] / (Co[a,a]+Co[b,b]-Co[a,b])$
- Salton (cosine)  $S[a,b] = Co[a,b] / \sqrt{(Co[a,a].Co[b,b])}$

In our analysis, we will use another normalization – activity or Balassa normalization.

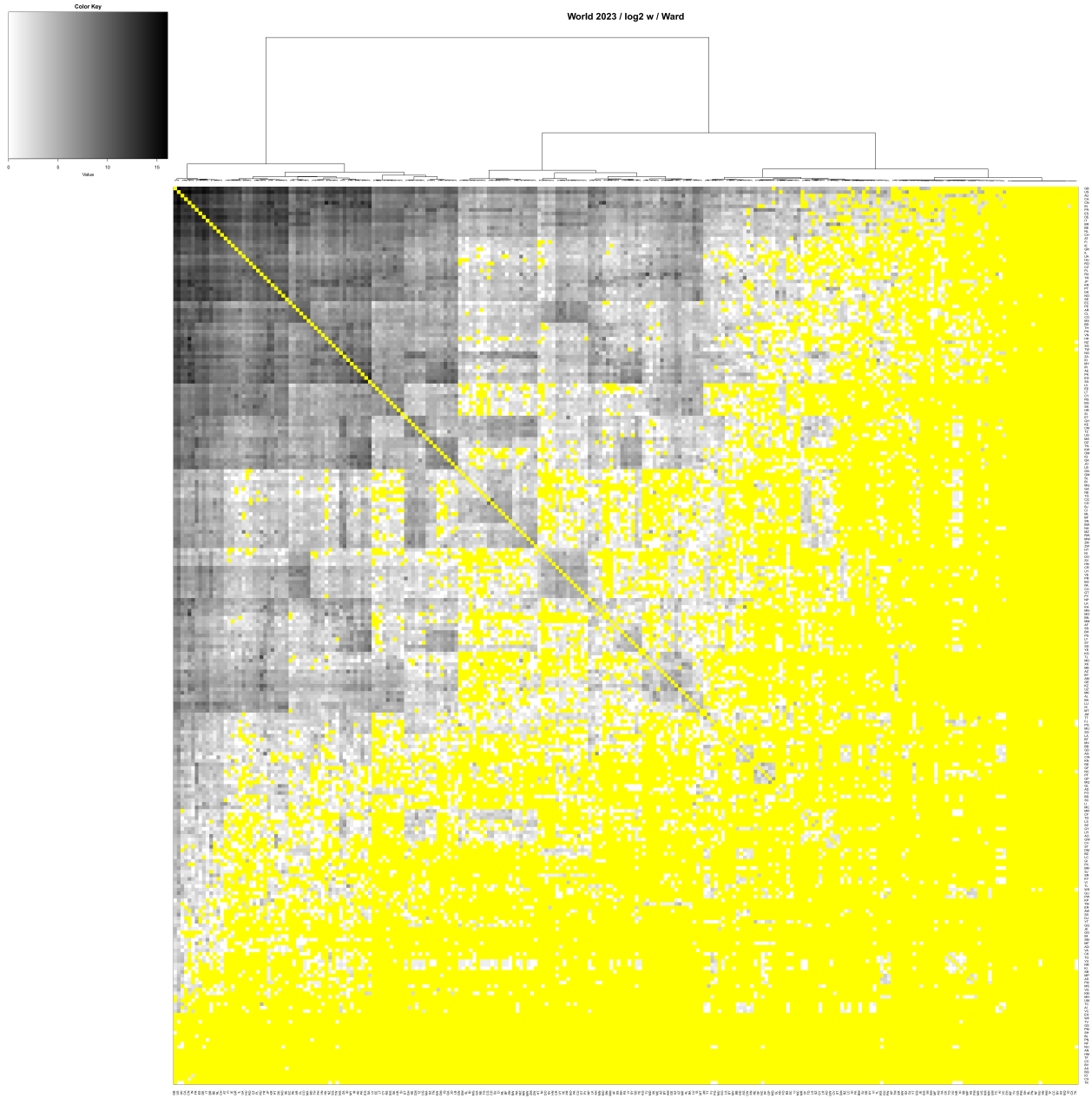

*Figure 2: Clustering of the co-authorship intensity matrix between world countries for 2023*

Let $Q(a)$ = windeg($a$) = $\sum_b Co[b,a]$ denote the column sum for the country $a$, and $T = \sum_{a,b} Co[a,b]$ the total sum of weights in the network. In our network $R(a) = Q(a)$. Then $R(a)/T$ is the probability of activity of country $a$. The expected weight E[$a,b$] from $a$ to $b$ is equal to:

$$E[a,b] = R(a) \cdot Q(b) / T$$

The measured weight $C[a,b]$ may deviate by a factor $A(a,b)$ from the expected value, $C[a,b] = A(a,b) \cdot E[a,b]$, or (Zitt et al., 2000, p. 633)

$$A(a,b) = C[a,b] \cdot T / (R(a) \cdot Q(b))$$

If $A(a,b) > 1$ the measured weight is larger than expected. The deviation measure $A$ is called the **activity** index (also the **Balassa** index or the "revealed comparative advantage" (Balassa, 1965)).

The range of *A* is not 'symmetric'. To symmetrize it, we apply a logarithmic function to it (Vollrath, 1991)). For easier interpretation, we selected base 2 logarithms:

$B(a,b) = \log_2 A(a,b)$, for $A(a,b) > 0$

If $B(a,b) = 0$, the collaboration is equal to the expected value. In our analysis, we used the index *B*. We have $A(a,b) = 0$ for non-linked countries. We set $B(a,b) = 0$ in such cases.

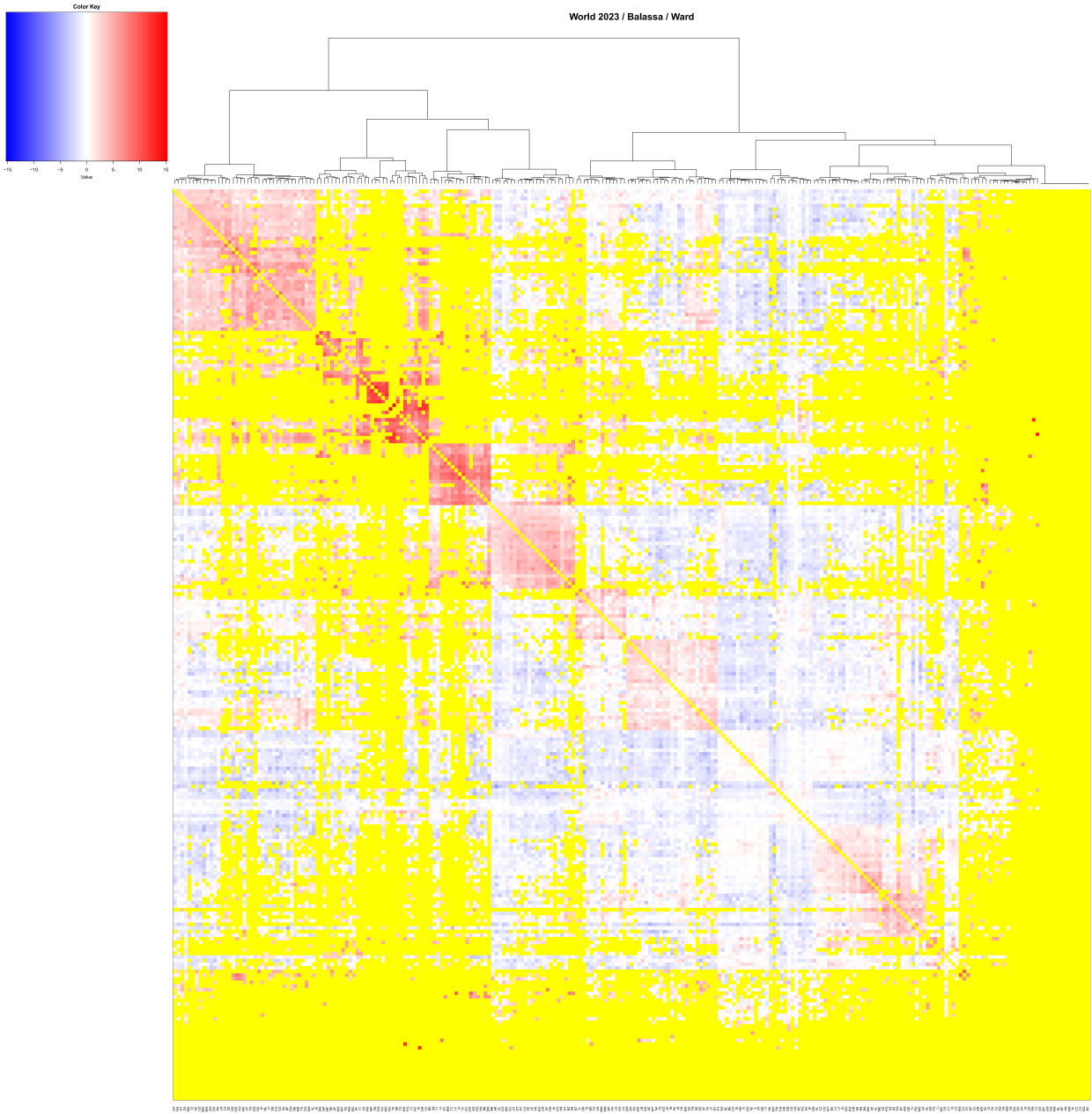

*Figure 3: Clustering of the Balassa index matrix for world countries co-authorship in 2023*

$B_1$ = { GH, NG, ET, ZA, RW, TZ, KE, UG, MW, BW, ZM, ZW, NA, LR, LS, SZ },
$B_2$ = { GN, GW, AO, MZ, CF, TD, GQ, GA, BI, ML, CI, SN, CG, CD, BJ, BF, CM, NE, MR, TG, SO, GM, SL },
$B_3$ = { JE, MS, GP, MF, RE, GF, MQ, MV, SC, MG, MU, PF, YT, CK, PW, MP, AS, FM, GU, MH, NU, TK, SB, TO, WS, PG, FJ, NC, KI, NR, VU },

$B_4$ = { BS, JM, TT, VI, KY, BM, TC, LC, AI, VC, GY, CW, AG, GD, KN, BB, DM },
$B_5$ = { BR, AR, CL, CO, MX, EC, UY, GT, PY, CU, PE, VE, PA, BO, CR, SV, HN, NI, DO, PR, HT, BZ, SR },
$B_6$ = { BT, TL, BN, ID, BD, TH, VN, MM, MN, NP, PH, LK, KH, LA },
$B_7$ = { OM, BH, PS, KW, LB, QA, AE, MY, SA, IQ, EG, JO, IN, PK, IR, TR, MA, DZ, TN, SS, SD, AF, YE, LY, SY },
$B_8$ = { PT, ES, NL, BE, CH, IE, DK, FI, NO, SE, IL, AT, DE, IT },
$B_9$ = { HK, MO, TW, CN, KR, GB, CA, US, NZ, SG, AU, JP },
$B_{10}$ = { GR, PL, CZ, HU, MT, EE, LV, LT, SI, RO, BG, HR, SK, RS, BA, MK, XK, AL, ME },
$B_{11}$ = { UZ, KG, AZ, KZ, TM, BY, AM, GE, TJ, RU, MD, UA },
$B_{12}$ = { GL, AD, FO, LI, MC, FR, CY, IS, LU },
$B_{13}$ = { KM, CV, ST, KP, DJ, ER, AW, SX, GI, FK, SJ, AQ, VA, SM, IM, GG, SH, AX, IO, VG, PN, TV },
$B_{14}$ = { EH, WF, UM, GS, PM, BL, NF, AN, HM, TF, CC, CX, BQ, BV }.

The diagonal blocks are mostly red or at least white – the inside cluster activity is larger than expected. The activity between most of the African countries from $B_1$ and $B_2$ is intensified. A very strong activity is between Pacific islands {PW, MP, AS, FM, GU, MH} and also inside the cluster of Caribbean islands $B_4$. The almost white diagonal blocks on the West European countries $B_8$ and other developed countries $B_9$ tell us that their collaboration is as expected. So is the collaboration between the West European countries $B_8$ and the East European countries $B_{10}$. Most of the out diagonal blocks are blue – less active than expected. An exemption is the block ($B_1$-$B_4$, {MA, DZ, TN, SS, SD, AF, YE, LY, SY}). There are some isolated dark red cells such as (WS, PU), (KI, TV), and (LC, AW) and some small red blocks such as ({GN, GW, AO, MZ}, {CV, ST}) and ({AW, SK}, {BB, AG, CW}). There are also some blue "lines" – mostly noncollaborative countries such as MO, HK, TJ, and SA.

## Conclusions

The article presents the first version of the R package OpenAlex2Pajek. There are some improvements planned. First, we will try to do the entire conversion in R. We will also expand the range of acquired data units and program a version of the package that performs the conversion from a local copy of the database.

OpenAlex is a rich source of bibliographic data relatively easy to use also from user's programs so enabling more demanding analyzes of bibliographic data. Here, it is important to ensure high data quality (Besançon et al., 2024), and OpenAlex users can play a big role with their feedback.

**Acknowledgments**

The computational work reported in this paper was performed using a collection of R functions OpenAlex2Pajek and the program Pajek for analysis of large networks. The code, data and color figures in PDF are available at Github/Bavla/OpenAlex.

This work is supported in part by the Slovenian Research Agency (research program P1-0294, research program CogniCom (0013103) at the University of Primorska, and research projects J5-2557 and J5-4596), and prepared within the framework of the COST action CA21163 (HiTEc).